\documentclass[12pt]{iopart}
\usepackage{iopams}
\usepackage{graphicx}

\begin{document}

\title{Influence of zonal flows on unstable drift modes in ETG turbulence}

\author{V M Lashkin$^1$, Yu A Zaliznyak$^1$ and A I Yakimenko$^{1,2}$}

\address{$^1$Institute for Nuclear Research, Kiev 03680, Ukraine \\
$^2$Department of Physics, Taras Shevchenko National University,
Kiev 03022, Ukraine} \ead{vlashkin@kinr.kiev.ua}

\begin{abstract}
The linear instability of the electron temperature gradient (ETG)
driven modes in the presence of zonal flows is investigated.
Random and deterministic $cos$ - like profiles of the zonal flow
are considered. It is shown that the presence of shearing by zonal
flows can stabilize the linear instability of  ETG drift modes.
\end{abstract}


\section{Introduction}

The electron temperature gradient (ETG) driven mode \cite {Horton,
Lee, Li} is often considered as a possible candidate for the
explanation of electron thermal transport through internal
transport barriers when ion temperature gradient (ITG) turbulent
fluctuations are suppressed by $\mathbf{E}\times \mathbf{B}$ shear
flow. At the same time, small scale ETG fluctuations have the
typical spatial scale of the order of electron gyroradius
$\rho_{e}$, and are less susceptible to quenching by shearing
\cite{Jenko, Gurcan1, Gurcan3}. It is widely thought that
drift-wave-type turbulence can excite zonal flows which are
associated with azimuthally symmetric band-like shear flows that
depend only on the radial coordinate \cite{Diamond05,DiamondRev2}.
Zonal flows play a crucial role in regulating drift-wave
turbulence and transport in tokamaks. It is now quite clear that
zonal flows are generated by modulational instability of drift
waves \cite{Smol1, Smol2, Chen, Manfredi,Zonca}. For ETG driven
turbulence, the excitation of zonal flow was considered in
\cite{HollandDiamond} (the case of broad turbulent spectrum) and
\cite{Zalik} (the four-wave coupling scheme).

It is well known that the presence of shear flow give rise not
only to instability of the sheared layer (Kelvin--Helmholtz
instability), but also to stabilization of other instabilities
(ITG driven modes, resistive interchange modes etc.)
\cite{Diamond05, Tajima, Hamaguchi, Waltz, Sugama}. Up to now, the
stabilizing effect on the linear instabilities has been considered
only for the case of mean smooth flows. Note, that zonal flow
shearing differs from that for mean flow shearing on account of
the complexity of the flow pattern. In contrast to smooth, static
mean flows, the zonal flow patterns can be expected to have finite
correlation time and complex, possibly random, spatial structure
\cite{Diamond05,Kim}.

In the present work we consider the influence of zonal flow on the
linear stability of ETG drift modes. We show that the presence of
random shearing by zonal flows strongly affects the linear
stability of ETG modes and has a stabilizing effect. If the mean
square amplitude of zonal flow exceeds some critical value, the
linear instability of ETG modes is suppressed for all poloidal
wave numbers $k_{y}$.

\section{Basic equations}

Assuming a slab two-dimensional geometry, charge quasineutrality
and the adiabatic ion responce, we consider the following
simplified model describing curvature driven ETG turbulence and
including viscosity and thermal diffusivity \cite{Gurcan1}
\begin{equation}
\label{eq1} \left(\frac{\partial}{\partial t}+\mathbf{v_{E}}\cdot
\nabla\right)(\varphi-\Delta_\perp
\varphi)+\frac{\partial}{\partial
y}(\varphi+p)+\nu\Delta^2\varphi=0,
\end{equation}
\begin{equation}
\label{eq2} \left(\frac{\partial}{\partial t}+\mathbf{v_{E}}\cdot
\nabla\right)p-r\frac{\partial}{\partial y}\varphi-\chi\Delta p=0,
\end{equation}
where $\mathbf{v_{E}}=[\mathbf{\hat{z}}\times \nabla\varphi]$ is
$\mathbf{E}\times \mathbf{B}$ drift velocity, $\varphi$ and $P$
are the normalized electrostatic potential and plasma pressure
respectively, and $\{A, B\} =
\partial_xA\partial_yB - \partial_yA\partial_xB$ is the Jakobian.
The variables in equations (\ref{eq1}) and (\ref{eq2}) have been
rescaled as follows
\begin{displaymath}
    r = \frac{\epsilon_B\epsilon_{*e}}{\epsilon_{*i}^2},
\end{displaymath}
\begin{displaymath}
 \varphi = \frac{1}{\epsilon_{*i}}\frac{e\phi}{T_i}, \qquad P =
 \frac{\epsilon_B}{\epsilon_{*i}^2}\frac{P}{P_{i0}},
\end{displaymath}
\begin{displaymath}
 x = \frac{x'}{\rho_{s}\sqrt{\tau}}, \qquad
 y = \frac{y'}{\rho_{s}\sqrt{\tau}}, \qquad
 t = \epsilon_{*i}\omega_{Bi} t',
\end{displaymath}
$x'$, $y'$ and $t'$ being the original physical coordinates (with
$x'$ the poloidal and $y'$ the radial coordinate),
\begin{displaymath}
 \epsilon_{*i} = \frac{\rho_s \sqrt{\tau}}{L_n}, \qquad
 \epsilon_B = \frac{\rho_s \sqrt{\tau}}{L_B}, \qquad
 \epsilon_{*e} = \frac{\rho_s \sqrt{\tau}}{L_p},
\end{displaymath}
where $L_n$, $L_B$ and $L_p$ are the background gradient scales
for the density, magnetic field and pressure respectively,
$\rho_s$ is the ion gyroradius calculated at the electron
temperature $T_{e}$, and $\tau = T_i/T_e$. The effect of zonal
flow can be included by assuming the $\mathbf{E}\times \mathbf{B}$
drift velocity of
\begin{equation}
\label{flow}
\mathbf{v_{E}}=v(x)\mathbf{\hat{y}}+[\mathbf{\hat{z}}\times
\nabla\varphi],
\end{equation}
where the first term accounts for zonal flow. Neglecting nonlinear
terms in equations (\ref{eq1}) and (\ref{eq2}), taking into
account equation (\ref{flow}) and representing
$$\varphi(x,y,t)=\Phi(x)\exp(ik_yy-i\omega t),$$
$$p(x,y,t)=P(x)\exp(ik_yy-i\omega t),$$
 one can obtain
\begin{equation}
\fl (-i\omega+ik_yv)(1+k_y^2-\frac{d^2}{dx^2})\Phi+ik_y(\Phi+P)
+\nu(k_y^4+\frac{d^4}{dx^4}-2k_y^2\frac{d^2}{dx^2})\Phi=0,
\label{e1}
\end{equation}
\begin{equation}
\fl (-i\omega+ik_yv)P-ik_yr\Phi+\chi(-k_y^2+\frac{d^2}{dx^2})P=0.
\label{e2}
\end{equation}
In the inviscid limit and absence of zonal flow, after taking
$\Phi(x), P(x)\sim \exp (ik_{x}x)$, equations (\ref{e1}) and
(\ref{e2}) give the dispersion relation for ETG modes
\begin{equation}
  \label{LinearDispersion}
    \omega_{1,2} = \frac{k_y}{2\left(k^2+1\right)}
     \left[1\pm\sqrt{1-4r\left(k^2+1\right)}\right],
\end{equation}
where $k^{2}=k_{x}^{2}+k_{y}^{2}$, and the plus sign describes the
drift waves dispersion, while the minus sign corresponds to the
dispersion of the convective cells. Equation
(\ref{LinearDispersion}) predicts instability with the growth rate
\begin{equation}
\label{growth}
\gamma=\frac{k_{y}}{2(k^{2}+1)}\sqrt{4r(k^{2}+1)-1},
\end{equation}
if $4r\left(k^2+1\right)\ge 1$.

\section{ETG drift modes in the presence of zonal flow}

We assume that the profile of zonal flow $v(x)$ is the random
function and $v(x)$ is a zero mean, $\langle v(x) \rangle =0$,
 real homogeneous Gaussian process with correlation function
\begin{equation}
\label{random} \left\langle v(x)v(x')\right\rangle=D(x-x'),
\end{equation}
where $\langle\ldots\rangle$ means statistical averaging. It is
assumed, that the intensity of the noise is small, $D(x)\ll 1$.
When neglecting $v^{2}$ term, and in the inviscid limit, equations
(\ref{e1}) and (\ref{e2}) reduce to the equation
\begin{equation}
\fl \omega^{2}\left(\frac{d^{2}}{dx^{2}}-k_{y}^{2}-1\right)\Phi+
\omega\left\{k_{y}-2k_{y}v\left(\frac{d^{2}}{dx^{2}}-k_{y}^{2}-1\right)\right\}\Phi
-k_{y}^{2}(v+r)\Phi=0,
\end{equation}
which after the Fourier transforming ($d/dx\rightarrow -iq$) can
be rewritten as
\begin{eqnarray}
\fl G_{0}^{-1}(p)\Phi(p)  +\int k_{y}[k_{y}-2\omega
(q_{2}^{2}+k_{y}^{2}+1)]v(q_{1})\Phi(\omega,q_{2},k_{y})\delta
(q-q_{1}-q_{2})dq_{1}dq_{2}=0, \label{main3}
\end{eqnarray}
where $p\equiv(\omega,q,k_{y})$ and
\begin{equation}
G_{0}^{-1}(p)=\omega^{2}(q^{2}+k_{y}^{2}+1)-\omega
k_{y}+k_{y}^{2}r.
\end{equation}
To avoid the cumbersome expressions, we write equation
(\ref{main3}) in the symbolic form
\begin{equation}
\label{main4}
G_{0}^{-1}(p)\Phi(p)+\gamma(p,p_{1},p_{2})v(p_{1})\Phi(p_{2})=0,
\end{equation}
where \begin{equation} \label{gamma}
\gamma(p,p_{1},p_{2})=\varkappa
(p_{1},p_{2})\delta(p-p_{1}-p_{2}),
 \end{equation}
 and integration over
repeated indices is assumed, as usual. Introducing a source $\eta
(p)$ in the right hand side of equation (\ref{main4}) and taking
the functional derivative $\delta/\delta\eta (p')$, we have
\begin{equation}
G_{0}^{-1}(p)G(p,p')+\gamma(p,p_{1},p_{2})v(p_{1})G(p_{2},p')=\delta
(p-p'), \label{main5}
\end{equation}
where $G(p,p')=\delta \Phi (p)/\delta\eta (p')$ is the Green
function, and we have taken into account that $\delta\eta
(p)/\delta\eta (p')=\delta (p-p')$. The function $G_{0}(p,p')$ is
a free (i. e. in the absence of the noise $v$) Green function.
Representing the Green function as a sum of the average and
fluctuating parts
\begin{equation}
G(p,p')=\langle G(p,p') \rangle+\tilde{G}(p,p'),
\end{equation}
substituting it into equation (\ref{main5}) and averaging, one can
obtain
\begin{equation}
G_{0}^{-1}(p)\langle G(p,p')\rangle+ \gamma(p,p_{1},p_{2})\langle
v(p_{1})\tilde{G}(p_{2},p')\rangle=\delta (p-p'), \label{main6}
\end{equation}
Substracting equation (\ref{main6}) from equation (\ref{main5}),
we get
\begin{eqnarray}
G_{0}^{-1}(p)\tilde{G}(p,p')+ \gamma(p,p_{1},p_{2})
v(p_{1})\langle G(p_{2},p')\rangle \nonumber \\ +
 \gamma(p,p_{1},p_{2})[v(p_{1})\tilde{G}(p_{2},p')-
 \langle v(p_{1})\tilde{G}(p_{2},p')\rangle]=0.
\label{main7}
\end{eqnarray}
In the Bourrett approximation \cite{Krommes}, which is justified
when the intensity of the noise is small enough, we can neglect
the term $v\tilde{G}-\langle v\tilde{G} \rangle$ and get for the
fluctuating part of the Green function
\begin{equation}
\tilde{G}(p,p')=-G_{0}(p)\gamma(p,p_{1},p_{2}) v(p_{1})\langle
G(p_{2},p')\rangle.
\end{equation}
Inserting this expression into equation (\ref{main6}), we obtain
\begin{equation}
\fl G_{0}^{-1}(p)\langle G(p,p')\rangle-
\gamma(p,p_{1},p_{2})\gamma(p_{1}',p_{2}',p_{2})G_{0}(p_{2})\langle
v(p_{1})v(p_{1}')\rangle\langle G(p_{2}',p')\rangle=\delta (p-p').
\label{main8}
\end{equation}
As follows from equation (\ref{random}), in the wave number domain
the correlator  has the form $\langle v(p)v(p')\rangle =I(p)\delta
(p+p')$. Due to homogeneity of the random process, the Green
function in equation (\ref{main8}) has the structure $\langle
G(p,p')\rangle=G(p)\delta (p-p')$. Making use of this and equation
(\ref{gamma}), one can perform some integrations in equation
(\ref{main8}) and finally we get
\begin{equation}
 G(p)=\frac{1}{G_{0}^{-1}(p)-\varkappa (p-p_{1},p_{1})\varkappa
(p_{1}-p,p)G_{0}(p_{1})I(p-p_{1})}.
\end{equation}
The poles of the Green function $G(p)$ determine the spectrum of
elementary excitations and the corresponding dispersion relation
is
\begin{equation}
\label{disp}
 G_{0}^{-1}(p)-\varkappa (p-p_{1},p_{1})\varkappa
(p_{1}-p,p)G_{0}(p_{1})I(p-p_{1})=0.
\end{equation}
In the absence of zonal flow we have $G_{0}^{-1}(p)=0$ and recover
the previous result (\ref{LinearDispersion}) with $k_{x}\equiv q$.
In what follows we consider the case when the profile of the zonal
flow is described by the random function $v(x)$ which has the form
\begin{equation}
v(x)=v_{0}\cos(q_{0}x+\vartheta),
\end{equation}
where the random amplitude $v_{0}$ is a zero mean, normally
distributed value with variance $\sigma^{2}$, and the random phase
$\vartheta$ is uniformly distributed between $0$ and $2\pi$. The
correlation function  (\ref{random}) of such a process is
$D(x)=(\sigma^{2}/2)\cos(q_{0}x)$ or, in the wave number domain
\begin{equation}
\label{peak}
I(q)=\frac{\sigma^{2}}{4}[\delta(q-q_{0})+\delta(q+q_{0})].
\end{equation}
\begin{figure}
\includegraphics[width=5in]{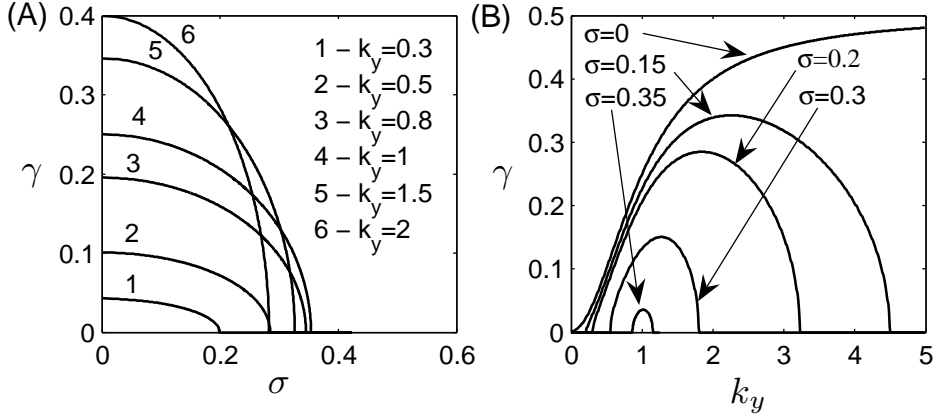}
\caption{(A) The instability growth rate as a function of the root
mean square zonal flow amplitude $\sigma$ (for different values of
the poloidal wave number $k_{y}$). (B) The growth rate as a
function of the poloidal wave number $k_{y}$ (for different values
of $\sigma$).} \label{fig1}
\end{figure}
In this case the noise has an infinite correlation length and is
concentrated at the wave number $q_{0}$ so that the characteristic
scale length of the zonal flow is of the order $\sim q_{0}^{-1}$.
As follows from eqution (\ref{growth}), in the absence of zonal
flow the maximum growth rate for the fixed $k_{y}$ is achieved at
$k_{x}=0$. Thus, to simplify calculations, we will consider the
influence of zonal flow on the most unstable modes and put $q=0$.
Then, substituting equation (\ref{peak}) into equation
(\ref{disp}) one can obtain the dispersion relation
\begin{equation}
\label{alg}
c_{1}\omega^{4}-c_{2}\omega^{3}+c_{3}\omega^{2}-c_{4}\omega+c_{5}=0,
\end{equation}
where
\begin{eqnarray}
c_{1}=(k_{y}^{2}+1)(q_{0}^{2}+k_{y}^{2}+1), \\
c_{2}=k_{y}(q_{0}^{2}+2k_{y}^{2}+2), \\
c_{3}=k_{y}^{2}(q_{0}^{2}+k_{y}^{2}+1)[r-2\sigma^{2}(k_{y}^{2}+1)]
+k_{y}^{2}[1+r(k_{y}^{2}+1)], \\
c_{4}=2k_{y}^{3}r-k_{y}^{3}\sigma^{2}(q_{0}^{2}+2k_{y}^{2}+2), \\
c_{5}=k_{y}^{4}\left(r^{2}-\frac{\sigma^{2}}{2}\right) .
\end{eqnarray}
Figure \ref{fig1}(a) shows the growth rate $\gamma$ of the ETG
driven mode as a function of the root mean square zonal flow
amplitude $\sigma$ for different values of the poloidal wave
number $k_{y}$. The characteristic wave number of zonal flow used
in the calculations is $q_{0}=0.1$. The parameter $r$ has been
fixed at $r=0.25$ so that, as follows from equation
(\ref{growth}), in the absence of random shearing there is a
linear instability for all $k_{y}\neq 0$. It is seen that the
growth rate decreases with increasing $\sigma$ and vanishes above
some value of $\sigma$ which depends on $k_{y}$. Thus, zonal flow
shearing stabilizes the instability of ETG modes. In figure
\ref{fig2}(b) we plot the dependence of the growth rate on the
poloidal wave number $k_{y}$ for different values of the root mean
square zonal flow amplitude $\sigma$. In the absence of zonal flow
($\sigma=0$) the growth rate $\gamma$ increases with increasing
$k_{y}$ and saturates at the level $\gamma_{max}=\sqrt{r}$. The
presence of shearing ($\sigma\neq 0$) changes the situation
drastically. The reduced growth rate initially increases as a
function of $k_{y}$ from some $k_{y,1}$ and then decreases,
becoming zero at some $k_{y,2}$. For not too large $\sigma$, the
linear instability is restricted to the region
$k_{y,1}<k_{y}<k_{y,2}$. The lower $k_{y,1}$ and upper $k_{y,2}$
boundary values increase and decrease respectively as the root
mean square zonal flow amplitude $\sigma$ increases and above some
critical value $\sigma_{cr}\sim 0.36$ (for $r=0.25$) the
instability  of ETG modes is suppressed for all poloidal wave
numbers $k_{y}$. An estimate for the critical value $\sigma_{cr}$
can be obtained from equation (\ref{alg}). The growth rate remains
zero as $k_{y}\rightarrow \infty$ (the most dangerous case) if
$\sigma=\sigma_{cr}$. From equation (\ref{alg}) one can see that
$\omega$ scales as $k_{y}^{-1}$ as $k_{y}\rightarrow \infty$.
Then, we can get the estimate $\sigma_{cr}=\sqrt{2}r$. This
theoretical prediction for the dependence of the critical value of
 root mean square zonal flow amplitude on $r$ is in very good
agreement with numerical results.

Next, we consider the case when the zonal flow profile is
deterministic and has the form $v(x)=v_{0}\cos (qx)$. In addition,
we include the effects of viscosity and thermal diffusivity. Then,
equations (\ref{e1}) and (\ref{e2}) can be rewritten as an
eigenvalue problem
\begin{equation}
\label{deter}
\left(%
\begin{array}{cc}
  k_y\hat{A}^{-1}\hat{B} &  k_y\hat{A}^{-1} \\
  -r k_y\hat{I} &   k_y\hat{C} \\
\end{array}%
\right)
\left(%
\begin{array}{c}
\Phi\\ P
\end{array}%
\right)=\omega\left(%
\begin{array}{c}
\Phi\\ P
\end{array}%
\right),
\end{equation}
where
\begin{eqnarray}
\hat{A}=1+k_y^2-\frac{d^2}{dx^2}, \quad \hat{B}=1+v_{0}\cos
(qx)\hat{A}-i\frac{\nu}{k_y}\hat{D}, \nonumber \\
\hat{D}=k_y^4-2k_y^2\frac{d^2}{dx^2}+\frac{d^4}{dx^4}, \quad
\hat{C}=v_{0}\cos (qx)-i\chi
k_y+i\frac{\chi}{k_y}\frac{d^2}{dx^2}. \nonumber
\end{eqnarray}
Employing a finite differencing approximation, we numerically
solved the eigenvalue problem (\ref{deter}). The dissipative
coefficients have been fixed at $\nu=\chi=0.01$. The instability
growth rate is plotted in figure \ref{fig2} for two values of the
zonal flow amplitude $v_{0}$ and different values of $q$ (for
$r=0.5$). It is seen that the presence of zonal flow reduces the
growth rate though the stabilizing effect manifests itself not so
sharply as in the case of random shearing.

\begin{figure}
\includegraphics[width=5in]{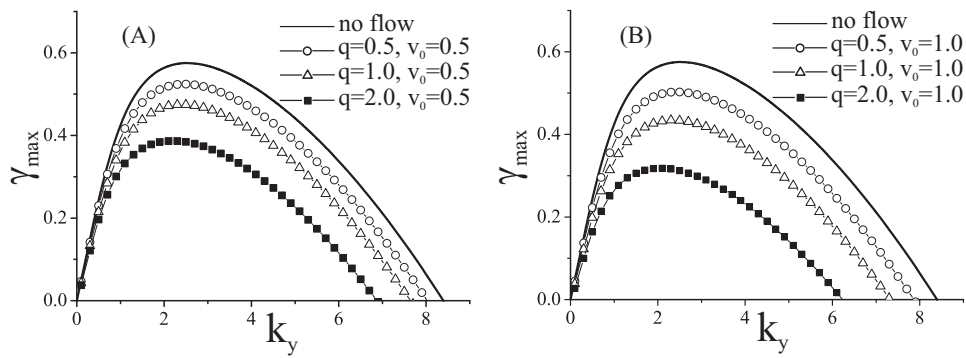} \caption{Maximum
growth rate versus $k_{y}$ for different values of $q$ in equation
(\ref{deter}) and (A) zonal flow amplitude $v_0=0.5$; (B) zonal
flow amplitude $v_0=1$.} \label{fig2}
\end{figure}

\section{Conclusion}

In conclusion, we have investigated the influence of zonal flows
on the linear instability of ETG driven modes. Random and
deterministic $\cos$ - like profiles of the zonal flow have been
considered. For the random profile of zonal flow, we have obtained
in the Bourrett approximation the dispersion relation for ETG
modes in the presence of shearing. We have shown that the presence
of random shearing caused by zonal flow has a strong stabilizing
effect on the ETG driven mode destabilized by the temperature and
pressure gradients. If the mean square amplitude of zonal flow
exceeds some critical value, the linear instability of ETG modes
is suppressed for all poloidal wave numbers $k_{y}$.

\section{Acknowledgment}
This work was supported by Ukranian Academy of Sciences through a
programme 'Fundamental problems in particle physics and nuclear
energy', grant number 150/166.

\end{document}